\documentclass[11pt]{article}

\usepackage{latexsym,amsmath,amssymb,theorem,epsfig} 
\usepackage{graphicx}

\DeclareFontFamily{OT1}{rsfs10}{}
\DeclareFontShape{OT1}{rsfs10}{m}{n}{ <-> rsfs10 }{}
\DeclareMathAlphabet{\mathscript}{OT1}{rsfs10}{m}{n}

\newcommand{\be}{\begin{equation}}
\newcommand{\ee}{\end{equation}}

\newcommand{\bea}{\begin{eqnarray}}
\newcommand{\eea}{\end{eqnarray}}
\newcommand{\ba}{\begin{array}}
\newcommand{\ea}{\end{array}}

\newcommand{\Tb}{\overline{T}}

\newcommand{\Zb}{\bar{Z}}

\newcommand{\Sb}{\overline{S}}

\def\IC{\mathbb{C}}

\newcommand{\beq}{\begin{equation}}

\newcommand{\eeq}{\end{equation}}

\newcommand{\ud}{\mathrm{d}}
\newcommand{\gen}[1]{\langle #1 \rangle}

\def\Tb{\bar{T}}
\def\Sb{\bar{S}}

\begin{document}

\begin{titlepage}
  \begin{flushright}
    hep-th/yymmnnn
  \end{flushright}
  \vspace*{\stretch{1}}
  \begin{center}
     \Large A New Method for Finding Vacua in String Phenomenology
  \end{center}
  \vspace*{\stretch{2}}
  \begin{center}
    \begin{minipage}{\textwidth}
      \begin{center}
        James Gray${}^{1*}$,
        Yang-Hui He ${}^{2,3\sharp}$,
        Anton Ilderton${}^{4\flat}$ and
        Andr\'e Lukas${}^{2\dagger}$
      \end{center}
    \end{minipage}
  \end{center}
  \vspace*{1mm}
  \begin{center}
    \begin{minipage}{\textwidth}
      \begin{center}
        ${}^1$Institut d'Astrophysique de Paris and APC, 
        Universit\'e de Paris 7,\\
        98 bis, Bd.~Arago 75014, Paris, France\\[0.2cm]
        ${}^2$Rudolf Peierls Centre for Theoretical Physics, 
        University of Oxford,\\ 
        1 Keble Road, Oxford OX1 3NP, UK\\[0.2cm]
	${}^3$Merton College, Oxford, OX1 4JD and
	  Mathematical Institute, Oxford University,\\
	  24-29 St.~Giles', Oxford OX1 3LB, UK\\[0.2cm]
        ${}^4$School of Mathematics and Statistics, University of Plymouth, \\
        Drake Circus, Plymouth PL4 8AA, UK
      \end{center}
    \end{minipage}
  \end{center}
  \vspace*{\stretch{1}}
  \begin{abstract}
    \normalsize 
    One of the central problems of string-phenomenology is to find
    stable vacua in the four dimensional effective theories which
    result from compactification.  We present an algorithmic method to
    find all of the vacua of any given string-phenomenological system
    in a huge class.  In particular, this paper reviews and then
    extends hep-th/0606122 to include various non-perturbative
    effects.  These include gaugino condensation and instantonic
    contributions to the superpotential.
  \end{abstract}
  \vspace*{\stretch{5}}
  \begin{minipage}{\textwidth}
    \underline{\hspace{5cm}}
    \\
    \footnotesize
    ${}^*$email: gray@iap.fr \\
    ${}^\sharp$email: yang-hui.he@merton.ox.ac.uk\\
    ${}^\flat$email: abilderton@plymouth.ac.uk\\
    ${}^\dagger$email: lukas@physics.ox.ac.uk 
  \end{minipage}
\end{titlepage}

\tableofcontents

\section{Introduction}

Much of the current effort within string phenomenology is directed
towards deriving effective theories with potentials capable of
stabilizing all of the moduli, see for example
\cite{Giddings:2001yu,Kachru:2003aw,Balasubramanian:2005zx,Grana:2005jc,Buchbinder:2003pi,Gray:2007mg,deCarlos:2004ci,Acharya:2006ia,Acharya:2007rc,Micu:2007rd}. In
some ways, the derivation of the four dimensional effective theory
relevant to the situation of interest is not the most difficult part
of such work. In many cases a mechanical framework exists, or is being
developed, that allows us to compactify a given higher dimensional
theory on one of the spaces of interest. There is not, however, an
equivalent mechanical framework which allows us to find the vacua of
the resulting four dimensional effective theory. Both the theory and
its vacua must be known before any phenomenological physics can be
extracted and so this represents a serious stumbling block,
particularly in studying the nature of spontaneous supersymmetry
breaking in these contexts. This paper is part of a series
\cite{Gray:2006gn} which attempts to address this problem by providing
a set of algorithmic methods for finding vacua of string derived or inspired four dimensional $\mathcal{N}=1$ effective supergravity theories\footnote{Related work, using similar
  techniques to investigate the vacuum geometry of gauge theories, can
  be found in \cite{probe}. Work on similar subjects employing
  algebro-geometric methods can be found in \cite{Distler:2005hi}.}.

The starting point for our analysis is such an ${\cal N}=1$ supersymmetric
theory, and in particular the associated K\"ahler and
superpotentials. Given these structures we employ a series of useful
tools, developed within the fields of computational 
algebraic geometry and
commutative algebra, to simplify the equations for the extrema of the
potential of the system. The basic idea is the following:
one of the reasons
that these equations are so complicated is that they describe many
different loci of extrema - an isolated minimum here, a line of maxima
there and so forth. The methods we discuss split these equations up
into a series of simpler systems - one describing each locus of
extrema. Further algorithmic methods then allow us to identify the
equations describing the loci of interest and to extract the physical
properties of the associated vacua.

In the first paper of this series, \cite{Gray:2006gn}, we applied our
methods to systems with superpotentials generated by perturbative
effects such as flux, torsion and non-geometric constructions. In the
present work we turn our attention to including non-perturbative
effects such as gaugino condensation and various instantonic
corrections. These contribute terms to the superpotential which are
exponential in the superfields. Naively, such transcendental behaviour
does not sit well with the methods of algebraic geometry which are
based upon the manipulation of polynomials. However, we shall show
that use of dummy variables to represent these exponentiated
quantities enables us to recover much of the power of our methods in
these cases. The methods we present will be of most use in dealing
with systems with both exponential and polynomial contributions to the
superpotentials.  The same techniques can be used to deal with
departures of the K\"ahler potential away from the pure sums of
logarithms considered in \cite{Gray:2006gn}. In particular, the
K\"ahler potentials which appear at conifold points in Calabi-Yau
compactifications, which contain logarithms of logarithms
\cite{Strominger:1995cz}, can be dealt with.

In addition to aiding in the finding of vacua and their properties, we
provide algorithmic methods for producing constraints on the
parameters of a system such that vacua of a desired type exist.
While not a complete cure for all problems in finding vacua, the
techniques which we present in this paper are a practical tool and in
very many systems of interest represent the best that one can do -
providing results which simply cannot be obtained by any other
existing method. The mathematics upon which the work is based also
represents perhaps the best way to formulate the problem of finding
vacua in these contexts.

The layout of this paper is as follows. In the next section we
describe our methodology. We begin with a brief recapitulation of the
methods, for finding perturbative vacua, presented in
\cite{Gray:2006gn}. We then move on to describe, in general, how to
adapt these methods to the case where non-perturbative effects are
included. In Section 3 we present the application of our method to
some examples taken from the string phenomenology literature. In
Section 4 we describe how to generate constraints upon the parameters
of a system which are necessary but not sufficient for the existence
of vacua of a desired type. Finally in Section 5 we briefly
conclude. The reader who is interested in the workings of the
algorithms mentioned in this paper, as opposed to simply what they do,
is referred to the appendices of \cite{Gray:2006gn} where a detailed
explanation is provided.


\section{The General Method} 
\label{generalmethod}
In this section, we describe in general the method which we will
employ to find stabilized vacua in flux systems. We will focus, in
particular, on how the problem can be phrased in terms of
computational algebraic geometry and be attacked algorithmically. We
devote the first subsection to a recapitulation of the methods of
\cite{Gray:2006gn} which studies the case of perturbative, polynomial,
superpotentials. There, standard methods of algebraic geometry
suffice. Next, we will examine how the presence of non-perturbative,
non-polynomial pieces in the superpotential introduces a
transcendental twist into the problem which, at first glance, seems
beyond the reach of commutative algebra. Nevertheless, we will devise
a systematic method wherein the variables which appear
transcendentally may be eliminated and the methods of the first
subsection once more become fully applicable. The same techniques may
be used to deal with K\"ahler potentials which differ in form from a
simple sum of logarithms of the real parts of the superfields. Such
theories are found, for example, 
in conifold limits of Calabi-Yau compactifications
\cite{Strominger:1995cz}.
\subsection{Finding Vacua for Perturbative
  Superpotentials}\label{s:pert}
In \cite{Gray:2006gn} we advocated, given the difficulty of finding
vacua of supergravities obtained by flux compactification, an
efficient and algorithmic methodology. In the absence of
non-perturbative terms, which generically contribute exponential terms
to the superpotential, we are typically confronted with an
extremization problem in complicated multi-variate polynomials.

In a typical example taken from the literature we have a K\"ahler
potential which is given by a sum of logarithms,
\begin{eqnarray} \label{K}
  K &=& - \sum_i b_i \log(M_i+\bar{M}_i).
\end{eqnarray} 
Here the $M_i=m_i + i \mu_i$ represent the moduli with $m_i$ and
$\mu_i$ being the real component fields. In what follows we will write
$q_I=\{m_i,\mu_i \}$ for the set of all real fields.  Our methods in
fact apply more generally and $K$ can be taken to be the logarithm of
an arbitrary polynomial of the complex fields and their conjugates. In
the interests of simplicity of presentation, however, we shall
restrict our discussion to K\"ahler potentials of the form
\eqref{K} in what follows.

The superpotential $W$ seen in flux compactifications is, in the
absence of non-perturbative effects and in the large volume and
complex structure limits, a polynomial function in the variables.
\bea 
\label{W} W &=& P(M_i) \ .
\eea

Given $K$ and $W$, \eqref{K} and \eqref{W}, the potential of the
system of interest is given by the standard result
\begin{eqnarray}\label{potential} 
V = e^K \left[ {\cal K}^{i \bar{j}} D_i W D_{\bar{j}} \bar{W} - 3 
  |W|^2 \right] \;.
\end{eqnarray} 
Here, as usual, the $D_i$ represents the K\"ahler derivative
$\partial_i\cdot +\cdot \partial_i(K)$ and ${\cal K}^{i \bar{j}}$ is
the inverse of the field space metric ${\cal K}_{i \bar{j}}
= \partial_i \partial_{\bar{j}} K$.  With our assumptions on the forms
of $K$ and $W$, the potential $V$ is then a rational function - a
fraction formed from two polynomials. Expanding all of the fields into
real and imaginary parts, we are left with a potential $V$ which is a
rational function of, say $2 n$, real variables. This is the potential
whose vacua we wish to find. Physically, we are not interested in the
solutions to the extremization equations $\partial V =0$ which are
given by taking the denominator to infinity (here the partial
derivative is taken with respect to the fields). These correspond to
the infinite field runaways common to these models. Therefore, it
suffices to consider simply the numerators of the expressions
$\partial V$ and ask that they vanish. We will denote the sets of
numerators of the partial derivatives of $V$ with respect to the
fields as $\gen{\partial V}$.

As stated in the introduction, the equations $\gen{\partial V}$ are in
general extremely complicated for systems of this kind. The reason for
this is that these equations contain a lot of information - that
concerning the location of all extrema of the potential at finite
values of the fields. We are of course only interested in a very
specific type of extremum of the potential, namely stable vacua. Given
this, the goal of the methods we present is to break up the system of
equations $\gen{\partial V}=0$ into a large set of simpler
systems. Each of these simpler systems will describe just one locus of
turning points. We then provide further algorithmic methods which pick
out the systems describing the extrema of interest.

In order to accomplish this decomposition of the equations we first
map the problem to one in algebraic geometry/commutative algebra.
The first insight of \cite{Gray:2006gn} in this direction was to
recognise that by temporarily complexifying the real fields, the list
of polynomials $\gen{\partial V}$ can be regarded as an ideal in the
polynomial ring $\IC[m_1, \ldots, m_n, \mu_1,\ldots, \mu_n]$. The
associated variety to this ideal is simply the loci of extrema of the
potential in field space.

The advantage of phrasing the problem in these terms is that the
mathematicians have an existing set of tools for breaking up ideals
into smaller systems of polynomials - in other words, there exist
algorithms which perform precisely the operations we need to isolate
our stable vacua.

We proceed according to the following steps. The details of the
algorithms are explained in \cite{Gray:2006gn} and can be performed on
the free computer algebra packages \cite{mac,sing}.

\begin{enumerate} 
\item {\bf Saturation Decomposition} 

  The first operation which we apply to our polynomials $\gen{\partial
    V}$ is what was referred to in \cite{Gray:2006gn} as the
  saturation decomposition.  Geometrically, the saturation
  $(I:f^{\infty})$ of an ideal $I$ with respect to a polynomial $f$ is
  the subspace within the variety described by $I$ which has $f$
  non-vanishing. In other words, for us it will be the space of
  extrema for which some polynomial, $f$, is non-zero. We also note
  that if we add a polynomial, $h$, to our ideal, denoting this by
  $\gen{\partial V, h}$, then this corresponds geometrically to all of
  the extrema of the system for which $h=0$.

  Let $F_{j}:= D_j W $ be the F-flatness equations, where
  $j=1,\ldots,n$ runs over the fields $M_j$, and $f_J$, for $J=1,
  \ldots, 2n$, their real and imaginary parts. The saturation
  decomposition is then the following `splitting up' of the
  polynomials $\gen{\partial V}$ describing the extrema of our
  potential:

  \bea\label{satexp} \nonumber \gen{\partial V} &=& \gen{\partial
    V,f_1,f_2,...,f_{2n}} \cap \\ \nonumber &&\bigcap\limits_{i}
  \gen{\left(\gen{\partial
        V,f_1,f_2,\ldots,f_{I-1},f_{J+1},\ldots,f_{2n}}:f_I^{\infty}\right)}
  \cap \\ \nonumber &&\bigcap\limits_{i,j}
  \gen{\left(\left(\gen{\partial V,f_1,f_2, \ldots ,f_{I-1},f_{I+1},
          \ldots, f_{J-1},f_{J+1}, \ldots, f_{2n}} :
        f_I^{\infty}\right):f_J^{\infty}\right)} \cap \\ \nonumber
  && \vdots \\
  && \gen{\left(\left(...\left(\partial V:f_1^{\infty}\right) \ldots :
        f_{n-1}^{\infty}\right):f_{2n}^{\infty}\right)} \ .  \eea

  The `$=$' symbol here means that the set of systems of polynomials
  on the right describe the same extrema as the large single system of
  polynomials on the left.

  Note that in practical examples the saturation expansion should
  always be performed before attempting a primary decomposition, which
  we will describe in the next step. 
  This process speeds up the computation hugely and turns
  what would otherwise be a method which is too slow to be of use into
  a fast and practical tool. The knowledge of suitable divisors with
  which to perform such a decomposition is non-trivial. The
  fact that the F-terms constitute a suitable set of polynomials is
  the result of a beautiful interplay between the supersymmetry of
  these systems and the algebraic geometry of their varieties of
  vacua.

  In addition to being practically essential, the saturation
  decomposition splits up the extrema in a physically useful
  manner. The first ideal in the decomposition above describes the
  SUSY vacua, the last ideal describes the vacua with spontaneously
  broken SUSY in which none of the F-terms vanish, while the other
  terms represent all cases between these two extremes. Which F-terms
  vanish is a useful manner of characterising the physical nature of
  the supersymmetry breaking which a vacuum exhibits and so, if one
  requires a particular kind of breaking, one can henceforth simply
  concentrate on the terms of interest.

\item {\bf Primary Decomposition}

  We can decompose our sets of polynomial systems into even smaller
  pieces. Firstly, an ideal $I$ may contain more information than is
  physically relevant. For example it might have multiplicities in its
  roots which describe the same vacua multiple times. Taking the
  `radical' $\sqrt{I}$ resolves this issue, as the radical is the
  maximal ideal associated with the variety of $I$. This step is also
  necessary for correctly identifying isolated extrema as we will
  describe shortly.

  If an ideal (when working over the complex numbers) is to describe a
  single locus of extrema then it must be what is called a `primary
  ideal'. The process of splitting an ideal up into primary pieces is
  called primary decomposition. A primary decomposition of each of the
  terms in \eqref{satexp} then furnishes us with a large set of simple
  polynomial systems describing all of the different vacua of our
  system. This set is automatically catalogued according to how they
  break supersymmetry.

  The algorithms which are required to perform the saturation
  decomposition, take the radical and perform the following primary
  decompositions are available in existing implementations in
  \cite{mac,sing}. The authors plan to make available in the future a
  Mathematica package which calls programs such as \cite{mac,sing}
  automatically.  This will make it possible to utilise our methods
  without knowing anything beyond the standard use of this mainstream
  program.

\item {\bf Isolated Vacua}

  Having split up the equations describing our extrema into separate
  pieces describing each loci of turning points, we may now proceed to
  pick out those of interest.

  We denote by $\{ \mathcal{P}_1, \ldots ,\mathcal{P}_q \}$ the set of
  all primary ideals with a specific pattern of supersymmetry
  breaking, corresponding to one of the terms in equation
  (\ref{satexp}). The isolated extrema, including all of the fully
  stable minima, have the property $\dim(\mathcal{P}_i)=0$. That the
  reverse is also true is guaranteed by taking the radical in the
  previous step as this removes the possibility of obtaining varieties
  which are subvarieties of other parts of the primary
  decomposition. This is important as we may otherwise have been led
  to believe that a dimension zero ideal was isolated, when it was in
  fact embedded in a higher dimensional part of the vacuum space.

  There exist algorithms which determine the dimension of ideals, again
  implemented in \cite{mac,sing}. In general the dimension of a
  primary ideal gives the number of flat directions of the associated
  vacua. This is true despite the complexification of field space
  which has taken place - the reader is referred to \cite{Gray:2006gn}
  for more details.

\item {\bf Physical Vacua}

  At this stage we have a set of polynomial systems describing all of
  the isolated vacua with the desired form of supersymmetry
  breaking. These systems, however, may not have any real solutions
  (recall that physically the variables here are real
  fields). Fortunately we may use existing methods from real
  algorithmic algebraic geometry to find out if this is indeed the
  case.

  For ideals of dimension 0, we can find the number of real roots by
  using methods based on what are called Sturm queries. More details
  of how this is done can be found in \cite{Gray:2006gn}. In practice
  it is rarely necessary to apply these methods in cases taken from
  the string phenomenology literature. It is normally the case that,
  by the time we have broken our systems of polynomials up to such an
  extent that we have obtained zero dimensional primary ideals, the
  resulting equations can be solved trivially by inspection.
 
\item {\bf Properties of Vacua}  

  In fact, we can do more with Sturm queries than simply find the
  number of real roots of our 0 dimensional systems. We can ask about
  the signs of any number of rational functions on these roots as well
  (again see \cite{Gray:2006gn} for details).

  In particular we may, therefore, algorithmically determine the
  number of isolated vacua with positive values for all of the
  eigenvalues of the Hessian. This is to say that we may determine the
  number of minima or, if preferred, the number of stable vacua
  satisfying the Breitenlohner-Freedman bound
  \cite{Breitenlohner:1982bm}. If we define, \bea g :=
  \left(\frac{\partial^2 V(q)}{\partial q_I \partial q_J} -
    \frac{3}{2} V(q) {\cal G}_{I J} \right) \ , \eea where ${\cal
    G}_{I J}$ is the metric on field space, then a vacuum is stable if
  the eigenvalues of $g$, evaluated at that point in field space, are
  all non-negative.

  Many other properties may also be determined algorithmically
  including the sign of the classical cosmological constant, whether
  the vacuum appears in a well controlled region of field space and so
  forth. Having accomplished our main goal of finding all of the
  isolated vacua of interest we now stop our summary of the
  main points of \cite{Gray:2006gn}.
\end{enumerate}

As a useful aside from our main discussion, we note that techniques
from algorithmic algebraic geometry may also be used to obtain
constraints on the parameters of a string phenomenology system
necessary in order for vacua of a given type to exist. The idea is as
follows. One again starts with the polynomials $\gen{\partial_I V}$
but now considers this system as an ideal in $\IC[a_l,m_i,\mu_j]$
where the $a_l$ are the parameters describing the number of units of
flux and so on. In other words, we consider the same set of
polynomials as before but now consider the surface that their zero
loci produce in the space of both fields and parameters.

Geometrically we would like to take this surface and project it on to
the plane spanned by the parameters $a_l$ alone. This would give rise
to a locus in this plane which is described by a series of equations
in the parameters. These would then constitute the desired set of
constraints.

The algebraic, algorithmic procedure which corresponds to this
projection is called Gr\"obner basis elimination.  This method is akin
to Gaussian elimination for linear systems - see \cite{Gray:2006gn}
for a description of the algorithm. As with the other algorithms
required for this work existing realisations may be found in
\cite{mac,sing}. Worked examples of obtaining such constraints, taken
from the string literature, may be found in \cite{Gray:2006gn}.

\vspace{0.1cm}
 
We have described in this subsection a five-step process which
formulates an algorithmic and, as it turns out, very efficient and
computerizable methodology in treating the problem of finding flux
vacua. In what follows, we will adhere to the general philosophy of
this five-step procedure, modifying it where necessary to accommodate the
introduction of exponential functions due to non-perturbative
effects.
 
\subsection{Finding Vacua for Non-Perturbative Superpotentials} 

The methods of the previous subsection constitute a powerful tool in the
context of locating perturbative vacua in models of string
phenomenology. However, it is frequently the case that such models are
not stable perturbatively and that effects such as gaugino
condensation and various kinds of instanton correction must be
included. In such situations the K\"ahler and superpotential,
\eqref{K} and \eqref{W} no longer adequately describe the physics of
interest. We will now describe how our methods can still be applied
even in these cases. We shall concentrate, for concreteness, on the
procedure to adopt when fields appear exponentiated in the
superpotential, a very common situation. Generalisations of this to
other situations, such as logarithms of logarithms in the
K\"ahler potential are straightforward.

\vspace{0.1cm}
 
Consider, then, a system with a K\"ahler potential given, as before,
by a sum of logarithms of the following form, \bea \label{Knew} K = -
\sum_i b_i \log ( M_i + \bar{M}_i) - \sum_a c_a \log ( N_a +
\bar{N}_a) \; .  \eea The superpotential, however, will now be taken
to contain both polynomial and exponential pieces, \bea \label{Wnew} W
= P(M_i, N_a) + \sum_a A_a e^{-d^a N_a }. \eea In general discussions
of our method we refer to moduli which appear only polynomially as
$M_i$. We write $N_a$ for variables which appear in the superpotential
both polynomially and in exponentials. In \eqref{Knew} and
\eqref{Wnew}, $b_i$,$c_a$,$A_a$ and $d^a$ are arbitrary constants. As
before $P$ is an arbitrary polynomial in its arguments. This example
of a class of models includes a large fraction of those found in the
literature. For example, any large volume and complex structure limit
description of a string phenomenological model could be expected to
take this form.

As before we obtain the potential for this system by use of the
standard supergravity formula \eqref{potential}. We then obtain the
equations for the extremization of this potential and demand that
these equations are satisfied by the vanishing of the numerator of
$\partial V$ in order to avoid the uninteresting solutions provided by
runaways to large field values.

Having obtained these equations we find that, as expected, they are no
longer polynomials but, instead, contain exponential terms. We render
this system polynomial again by introducing dummy variables for the
exponential terms that appear.  We shall use as our redundant
variables the real and imaginary parts of the moduli $M_j$ and $N_a$
along with the real and complex dummy variables $x_a$ and $y_a$
defined as follows (with no implicit summation on the index $a$):
\begin{eqnarray} 
\label{newvar}
  M_j &=& m_j+i \mu_j \;\; , \;\; N_a=n_a+i\nu_a, \\ \nonumber
   x_a &=&\exp(-d^a n_a) \;\; , \;\;  
  y_a=\exp(-i d^a \nu_a). 
\end{eqnarray} 
As in the perturbative case we can now temporarily complexify the real
fields $m_j$, $\mu_j$, $n_a$, $\nu_a$, $x_a$ as well as making the
phase $y_a$ an arbitrary complex number. In this complexified
redundant field space one can now express the locus of extrema of the
potential as an algebraic variety and apply much of the methodology
outlined in the previous subsection. In particular, the first two
steps in our five point procedure, the saturation expansion and
primary decomposition, can now be applied. The only subtlety in doing
this is that the F-terms must also be written in terms of the
variables \eqref{newvar}.

We cannot, however, proceed with steps three through five. The problem
is that we cannot identify isolated vacua simply by demanding that
the dimensions of the primary ideals obtained after the first two
steps are zero. This is because, due to the variables \eqref{newvar}
being redundant, account must be taken of the relationship between
$x_a$ and $n_a$ and $y_a$ and $\nu_a$.

In order to circumvent this problem we must eliminate the hidden
transcendental nature of the system at this stage and the dummy
variables which arise. We shall do this by identifying all of
the $N_a$ values which can be associated with stable vacua. By
substituting the appropriate values for $n_a$, $\nu_a$, $x_a$ and
$y_a$ back into $\gen{\partial V}$ we shall arrive back at a set of
polynomial systems describing the vacua of interest which do not
involve dummy variables. We will then be able to continue with our
usual analysis. 

\subsubsection{Equations for the Transcendental Variables}
Let us discuss the decomposition and elimination of the transcendental
variables $N_a$ in detail. It is expedient to exemplify our methodology
with the pair of variables $(n_a,x_a) = (n,x)$ for some $a$. They are
related, by \eqref{newvar}, as \beq\label{x-n} x = e^{-d n} \; . \eeq
The imaginary part of $N_a$ and $y_a$ can be treated similarly.

We project one of the irreducible varieties, say $\mathcal{P}$, furnished by 
the first two steps of saturation and primary decomposition described
in Section \ref{s:pert}, to the $n$-$x$ plane. 
This will give us an ideal in $\IC[n,x]$ which we will refer to
as $\tilde{\mathcal{P}}$. This is achieved by performing an elimination-ordering
Gr\"obner basis calculation as described in \cite{Gray:2006gn}. This
gives us one of four possibilities. We will obtain either the full
plane as a variety, a complex curve in this plane, a point in the
plane or nothing at all.
 
In the case where there is nothing at all, $\tilde{\mathcal{P}}$ is
said to be of dimension $-1$ and there is no vacua within this term of
the decomposed saturation expansion.

The cases where the projected variety is a point
or a curve in the $x$-$n$ plane are more interesting. When combined
with \eqref{x-n} a set of possible solutions could give isolated
values for $n$. We must enumerate and find all of these possible
values for these fields (recall that we are interested, of course, in
real valued results). The idea is then to substitute each possible
combination of the values obtained for the set of fields $N_a$ back
into the original equations, and then analyze the resulting set of
varieties one by one to see if the other fields are similarly
stabilized. Once we have eliminated all of the $N_a$ we find ourselves
back in the case described in the previous subsection and our standard
procedure then applies.

The case where we obtain the full $n$-$x$ plane needs more
consideration.  Supposing that we have only one pair of $(n,x)$ type
variables, or more precisely only one pair of such variables where
both $n$ and $x$ appear in the equations, then obtaining the plane
also corresponds to physically undesirable results. We know that the
physically allowed values for $x$ and $n$ are related by \eqref{x-n};
this defines a (transcendental) curve in the plane.  This does not,
however, restrict the value of $n$ to a discrete set and so there is
an unstabilised flat direction of the extrema \footnote{The careful
  reader may be worried that these different possible values of $n$
  could be disconnected on the original variety and that the
  appearance of this ``flat direction'' might be an artifact of
  projecting down onto one plane. This objection is answered by first
  performing a primary decomposition on the ideal and then
  individually projecting each of the resulting irreducible
  varieties. This is related to a point to which we will shortly
  return and so we do not dwell on it here.}. Since we are interested
in completely stabilized vacua we will then discard these cases for
now.

The situation is different when we have multiple pairs of
$(n_a,x_a)$. It is sufficient to restrict attention to two pairs
$(n_1,x_1)$ and $(n_2,x_2)$. Projecting onto the $(n_1,x_1)$ plane
requires us to eliminate both of the variables $n_2$ and $x_2$. In the
elimination these are treated as independent variables, as there is no
way to impose the transcendental relationship between them in an
algebraic manner. This process may lead us, upon arriving at the whole
$(n_1,x_1)$ plane as a solution, to believe there is a flat direction
in field space, as in the previous paragraph. This is not necessarily
the case as we are missing some information - the transcendental
relationship between $n_2$ and $x_2$. One can not simply project down
the transcendental relation separately as the intersection of the
projection of two surfaces is not the same as the projection of their
intersection.

The resolution of this problem, in the case that one obtains the whole
plane for one or more projections, is to instead eliminate all of the
variables $m_a$ which enter only algebraically, and solve the
remaining transcendental equations numerically. This already presents
a significant simplification of the complete problem, and once this has
been performed one may return immediately to the perturbative stages
of the algorithm described earlier in the paper, as any transcendental
terms have been eliminated.

The examples in this paper will contain two pairs of variables which
appear transcendentally. However, one of these will be an axion and it
is a feature of the models we consider that, while the dummy variable
representing an exponentiated axion appears in our equations, the
axion itself does not. Indeed such a situation is very common in
string phenomenology. We normally only have to consider
non-perturbative effects if a modulus can not be stabilized
perturbatively - if, for example, the associated superfield can not
appear polynomially in the superpotential. Given such a situation, and
the fact that the K\"ahler potential often only depends upon the the
real parts of the superfields, we would discover that the exponential of
the axion appears in the ideals describing the extrema of the potential
but the field itself does not. Hence, when eliminating this variable
there is no missing information as the exponentiated field may itself
be taken to be a non-redundant description of the relevant degree of
freedom.

\vspace{0.1cm} 

\begin{figure}[!ht]
\centerline{\epsfxsize=4in\epsfbox{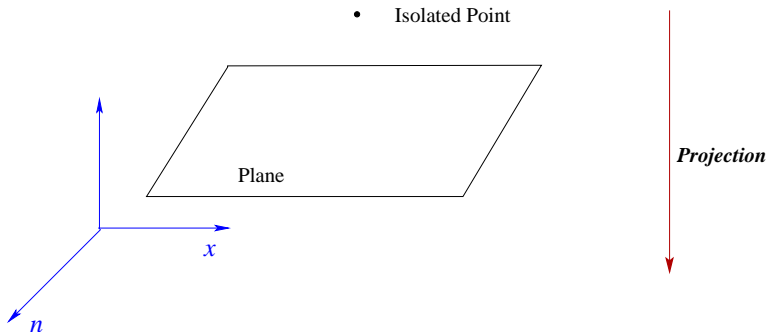}} 
\label{fig1}
{\it \caption{A reducible variety with a plane `obscuring' a point. In this
case, projection onto the $n$-$x$ plane will miss the isolated
point. However, primary decomposition will avoid this problem since
the point and the plane will be separated as different prime ideals.}}
\end{figure}

We remark that it is vital that the primary decomposition step be
performed before the elimination described in this subsection. If this
is not done it is possible that there could be isolated vacua which
are missed by the aforementioned elimination procedure. Imagine the
situation, depicted in Figure \ref{fig1}, where the variety associated
to $\mathcal{P}$ consists of two disconnected regions: a point and,
separated from this by some distance in the direction of projection, a
plane, which is not orthogonal to the $n$-$x$ plane upon which we
project.

\vspace{0.1cm}

If we were to encounter such a case in employing our method we would
erroneously conclude that the continuum of $n$ values obtained upon
projection means that there are no stabilized vacua in this
variety. Primary ideals, when we are working over a suitable
coefficient field, only include varieties of the same dimension and so
such ``obscuring'' of isolated vacua cannot occur if we only apply
the projection method in these cases.

\subsubsection{Solutions for the Transcendental Variables} 

We now have to solve the transcendental equations given by $\tilde{P}$
and \eqref{x-n}. In the examples we shall look at this reduces to the
study of a set of polynomials in $n$ and $e^{-d n}$ on the complex
$n$-plane. In general, finding the zeros of these functions will not
be analytically tractable~\footnote{In certain cases there are
  analytic functions, such as the Lambert W function, which do indeed
  solve the system. In these cases, we are studying the ring of
  polynomial (or the field of rational) functions extended by such
  transcendental functions, whereby furnishing an interesting
  generalisation of algebraic geometry.}. We shall therefore require
numerical methods at this stage. However, we wish to ensure that we do
not miss any solutions by employing such techniques. In order to
achieve this we count the number of solutions to our equations
analytically. We then simply hunt numerically for solutions until we
have a complete set.

Take one of the polynomial generators of $\tilde{\mathcal{P}}$ and
substitute in the defining equation \eqref{x-n} for $x$.  For such an
analytic function $f(n)$, we may enumerate the number of real roots (counted with multiplicities)
$Z_R[f]$ in a range, say $1<n<R$, by taking a contour integral,
\beq\label{roots} Z_R[f] = \frac{1}{2\pi i}\oint d
w\,\,\frac{f'(w)}{f(w)} \ , \eeq around the contour about the real
axis shown below:
\[ 
\includegraphics[width=0.7\textwidth]{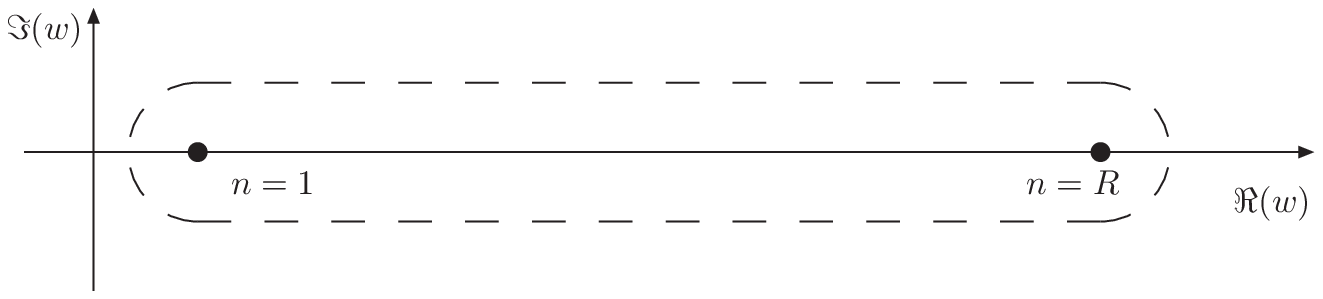} 
\] 
There is a risk of picking up complex roots to the equation but this
may be minimised by taking the width of the strip to be small. The
contour integral returns the number of real zeros of the function
$f(n)$ in $1<n<R$ counted with their multiplicities. If this is zero
then there are no solutions to our equations in the given region.
 
If, on the other hand, the integral is non-zero, we use a uni-variate
numerical root finding algorithm, such as those implemented in
\cite{mathematica,maple} to locate one root.  We then perform the
integral \eqref{roots}, now along a small circular contour about the
root in order to count its multiplicity. If this does not saturate
$Z_R[f]$ we return to the numerical root finder with different initial
values until we find a second root, count its multiplicity and repeat
until we have found all roots.  This process must be repeated for each
generator of the ideal and the list of roots compared to find which
are the simultaneous zeros. We must do so for all transcendental
variables $n_a$ and $\nu_a$. When all of the solutions have been found
for all of the variables, we substitute each possible combination of
values back into the original ideal to fully eliminate the
transcendental variables $N_a$. We are then left with a set of ideals
built out of the (real and imaginary parts of the) ordinary variables
$M_i$ and so algebraic methods suffice.

It is important to note that when substituting the obtained values for
$N_a$ back into the original ideal $\gen{\partial V}$ one must also drop
the equations for the extremization with respect to the $N_a$ as
generators. We have obtained the solutions to these equations
numerically and thus they are not exact. When substituted into
$\gen{\partial V}$ they will not, therefore, give rise to exactly
vanishing polynomials for values of the $M_i$ lying on the
variety. One would thus frequently conclude $\dim(\gen{\partial V})=-1$ and
that there are no vacua if one did not remember to drop these
equations once they have solved. The remaining generators of
$\gen{\partial V}$ can, of course, be derived from the original potential
with our values for the $N_a$ substituted into it. 

We are now in a position to return to standard methods as described in
Section \ref{s:pert} by forming the ideal of derivatives of $V$ with
respect to the remaining variables.  The new ideal may be written in
terms of polynomials of the real and imaginary parts of the remaining
moduli. The five-step procedure of the previous subsection can then be
applied to find all of the vacua of this system and their physical
properties.

In fact, one further check is required to avoid spurious results. It
is possible that, at the same $N_a$ value as a true isolated extremum,
there are points which extremize the potential with respect to the
$M_i$ but not the $N_a$. That we are not obtaining values of this
nature may simply be checked by ensuring that $\partial V / \partial
N_a =0$ for any vacuum found, to within the required numerical
accuracy. Since the relevant partial derivative is a rational function
of the $M_i$ when evaluated at our $N_a$ solution, and since any
decimal number of given precision may be approximated by a rational
number, this may be achieved within the algorithmic framework
advocated here.

\vspace{0.1cm}

In the presence of non-perturbative effects, then, the five-point
procedure of Section \ref{s:pert} should be replaced with the
following:

\begin{enumerate}
\item Introduce dummy variables for any exponentiated quantities;
\item Saturation and primary decomposition of the ideal
  $\gen{\partial V}$ expressed in terms of these redundant quantities;
\item Elimination of transcendental and dummy variables by the
  projection and numerical solution method described in this
  subsection;
\item Return to five-step procedure of Section \ref{s:pert} now that
non-polynomial properties of the system have been removed.
\end{enumerate}
 
\vspace{0.1cm}

Armed with this general method for finding the stabilised vacua in
systems with non-perturbative superpotential terms, we will now go on
to provide some examples of its application.
\section{Examples of Finding Supersymmetric and Non-Supersymmetric
  Vacua}
One could readily apply the technology described in section
\ref{generalmethod} to obtain many interesting results within the
context of flux vacua. In what follows we shall provide a variety of
examples such that the full power of our approach may be demonstrated.

\label{eg-het}
\subsection{A Heterotic Example}
Let us focus on a heterotic example taken from \cite{Micu} with
K\"ahler and super potentials as follows:
\begin{equation}\begin{split}\label{sat-W2}
    K&=-3\log(T+\Tb)-3\log(Z+\Zb)-\log(S+\Sb), \\
    W&=-80 T -36 i T Z + \frac{3}{2} T Z^2 + 480\, e^{-S/4} \ .
\end{split}\end{equation}
We now apply our method as described in Section
\ref{generalmethod}. We shall work our way through the algorithmic
process of finding vacua step by step, following the procedure given at
the end of that section.

\subsubsection*{Step 1: Introduction of Dummy Variables for
  Non-Polynomial Quantities}
In the notation of Section \ref{generalmethod}, $\{ M_i \} = \{ T,Z
\}$ and $\{ N \}= \{ S \}$. Furthermore, recalling \eqref{newvar}, we
define the component fields and dummy variables by 
\beq\label{newvar2}
T=t+i\tau, \ Z=z+i\zeta, \ S=s+i\sigma; \quad x = \exp(-s/4), \ y =
\exp(-i\sigma/4) \ .  
\eeq 
In what follows we write $f_1$, $f_2$ for
the real and imaginary parts of $F_T = D_T W$, $f_3$, $f_4$ for those
of $F_Z = D_Z W$ and $f_5$, $f_6$ for those of $F_S = D_S W$.

\subsubsection*{Step 2: Saturation and Primary Decomposition}

The next step is to perform the saturation decomposition on the ideal
$\gen{ \partial V}$ and primary decompose the resulting ideals.  The
table below displays the results of the saturation decomposition. In
fact, in forming this table we have jumped ahead slightly in that we
have only included those terms in the output of the decomposition
which include vacua (physical or otherwise). All other terms in the
saturation expansion are found, upon investigation, to correspond to
empty varieties.
\begin{center}
\begin{tabular}{|c|c|c|}\hline 
  Ideal & Vacua type & Physical Vacua?\\ \hline\hline
  $\langle f_1,f_2,f_3,f_4, f_5,f_6\rangle$ & supersymmetric & Yes, AdS 
  saddles \\ \hline\hline
  $(\gen{\partial V,f_1,f_2,f_4,f_5,f_6}:f_3^\infty)$ & partially F-flat & 
  No, 
  $z=0$ \\\hline
  $(\gen{\partial V,f_1,f_3,f_4,f_5,f_6}:f_2^\infty)$ & partially F-flat & 
  No, $t=0$ 
  \\\hline\hline
  $((\gen{\partial V,f_3,f_4,f_5,f_6}:f_1^\infty):f_2^\infty)$ & partially 
  F-flat & 
  No, $t=0$ 
  \\ \hline\hline
  $(((\gen{\partial
  V,\tau,f_2,f_4,f_6}:f^\infty_1):f^\infty_3):f^\infty_5)$ & non  
  SUSY & Yes, AdS saddles \\ \hline 
\end{tabular}
\end{center}
For the most computationally expensive parts of the saturation
decomposition, for example where all real and imaginary parts of the
F-terms are non-zero, we have restricted our attention to vacua where
some of the axions vanish. Such restrictions can easily be implemented
by simply adding the relevant fields to the list of generators of the
ideal being studied.

The supersymmetric variety in this system turns out to be an AdS
saddle point with a flat direction. Since this is not so interesting
let us demonstrate the rest of our method by focusing on the ideal
$(((\gen{\partial
  V,\tau,f_2,f_4,f_6}:f_1^\infty):f_3^\infty):f_5^\infty)$ in the last
row of the above table. This constitutes a nice example as, as we will
see, it will yield non-supersymmetric vacua.

After obtaining this ideal from the saturation decomposition we
primary decompose it and find \emph{two} pieces, let us call them
$\mathcal{P}_1$ and $\mathcal{P}_2$. These primary ideals constitute
10th order polynomial systems, each comprised of 27 generators; for
brevity we do not present them here. This ends the second step of our
procedure.

\subsubsection*{Step 3: Elimination of the Transcendental and Dummy Variables}
Taking one of the primary ideals from step 2, $\mathcal{P}_1$, we
project down onto the planes in field space spanned by an
exponentiated field and its associated dummy variable as described in
Section \ref{generalmethod}.

In this example, projecting onto the $( \sigma, y)$ plane, we find the
following ideal:
\bea \gen{ y-1} \ . \eea 
Meanwhile, projecting onto the $(s,
x)$ plane we find the ideal 
\bea &&\gen{7 s^{10}-197 s^9+2132 s^8-11744
  s^7+38464 s^6-69888 s^5-8192 s^4+16384 s^3 \\ \nonumber && -262144
  s^2-1114112 s+1835008} \ .  
\eea

Similarly, carrying out the same projections on the second primary
ideal $\mathcal{P}_2$, we find the following ideals:
\bea
&&\gen{y+1} \ , \\
&&\gen{7 s^{10}-197 s^9+2132 s^8-11744 s^7+38464 s^6-69888 s^5-8192
  s^4+16384 s^3 \\ \nonumber && -262144 s^2-1114112 s+1835008} \ .
\eea

It is a peculiarity of this particular example that the ideals found
in each case only depend on one of the variables. Note, in addition,
that both primary ideals lead to the same possible values for $s$.

We now come to the numerical part of our general method. Although we
have provided a completely general algorithmic discussion in this
paper for completeness and for ease of automation, we see that the
complex analysis aspects are not required for the present example.
This is because the saturation and primary decompositions prove to be
so powerful in breaking up the equations that one can solve the
resulting primary ideals trivially.

As we have already stated, the projected ideal in $(s,x)$ is a
polynomial in $s$ only, to which there are only two real and positive
solutions:
\begin{equation}
  s_1 = 1.21869\ldots \simeq \frac{574}{471},
  \quad s_2 = 9.68026\ldots \simeq\frac{4511}{466}.
\end{equation}
We have written decimal approximations to an error of within
$10^{-5}$ in the above expressions.
Corresponding values for the exponentiated values, $x$ can
then, of course, be also written down.

Examining the remaining ideals, $\gen{y-1}$ and $\gen{y+1}$, 
we find that the dilaton axion takes
the values $\sigma=4(2n\pi)$, $n\in\mathbb{Z}$ in the variety
corresponding to $\mathcal{P}_1$, and $\sigma=4(2n+1)\pi$ in the
variety of $\mathcal{P}_2$.

We thus have all of the possible values which the variables appearing
transcendentally in our equations can take in these vacua. This ends
step 3.

\subsubsection*{Step 4: Analysis of the Resulting Perturbative System}

The non-polynomial part of the problem has now been solved. By
substituting in all possible values for the exponentiated fields, as
obtained in step 3, into our expressions we can obtain a problem which
depends only upon variables which appear polynomially.

From Step3, we have \emph{four} sets of values for the dilaton and its axion
following from the two primary ideals (since $\sigma$ appears nowhere
except the exponential, the only variation between vacua caused by its
value in vacuum comes from the choice of even or odd multiples of
$4\pi$). Therefore we have four systems with perturbative potentials
to investigate. The scalar potential in these theories is given by one
of $V(T,Z,s_j+4i(2n\pi))$ or $V(T,Z,s_j+4i(2n+1)\pi)$ where $j=1$ or
$2$. We refer to these four with the notation $V_{j,\text{even}}$ and
$V_{j,\text{odd}}$ respectively. We now return to the five step
perturbative method as described in \cite{Gray:2006gn} and Section
\ref{s:pert}.

\subsubsection*{Saturation Decomposition and Primary Decomposition}
We take one of $V_{j,\text{even}}$ or $V_{j,\text{odd}}$ and calculate
the $T$, $Z$ derivatives of $V$. Still taking $\tau=0$ (as we had
above) we construct the ideal $\gen{\partial_T V,\partial_Z
  V,\tau}$. The resulting ideal is in fact so simple that there is no
need to perform a saturation decomposition with respect to the F-terms
of the full system.

Primary decomposing instead, we find that there is one primary ideal
for each of $V_{j,\text{even}}$ and $V_{j,\text{odd}}$, in the
remaining variables $t,z$ and $\zeta$.

\subsubsection*{Finding Isolated Vacua}
Using the technology of Sturm Queries, as described in
\cite{Gray:2006gn}, we find that each primary ideal here is zero
dimensional, corresponding to isolated vacua. For example, for
$V_{1,\text{odd}}$ this ideal is
\begin{equation}\begin{split}\label{ideal1odd}
    \langle&\zeta-12, \\
    &2397591226947591 z^8-115732161484426080 z^6-6077206297096777728 z^4 \\
    &-1267641623933014056960 z^2 +11126144838499707191296, \\
    &239281769562958390871733867 z^6+19000625879574432233531888112 z^4 \\
    &-4048205807673782004499951097088 z^2 +74098944342403413898555305459712 t \\
    &-360142886662687721358185689387008\rangle.
\end{split}\end{equation}

\subsubsection*{Finding Physical Vacua}

More Sturm Query methods allow us to ascertain that the ideals
corresponding to the $V_{j,\text{even}}$ have no real solutions, for
either $j=1$ or $2$, for which both $t$ and $z$ are
positive. Therefore all of the corresponding vacua are unphysical.

For $V_{1,\text{odd}}$, shown in \eqref{ideal1odd}, there are
physical solutions. One now has to check, as described in Section
\ref{generalmethod}, that the $S$ derivatives of the potential vanish
at these solutions. This is the case for a single solution which is
given below.
\begin{equation}\label{keepme}
  \zeta=12,\qquad t \simeq \frac{1489}{281},\qquad z\simeq \frac{762}{263}.
\end{equation}

For $V_{2,\text{odd}}$ there is again one physical solution,
\begin{equation} \label{keepme2}
	\zeta=12,\qquad t \simeq \frac{497}{314},\qquad z\simeq \frac{638}{347}.
\end{equation}

\subsubsection*{Properties of the Resulting Vacua}
The technology employing Sturm Queries can then be used to find out a
great deal of useful information about the resulting vacua, as
described in detail in \cite{Gray:2006gn}. For example, we find that
both of these solutions are isolated anti-de Sitter saddle points. The
vacuum shown in \eqref{keepme2} is Breitenlohner-Freedman stable while
the extremum given in \eqref{keepme} is not.

\begin{figure}[!!!t]
\centering\includegraphics[width=0.45\textwidth]{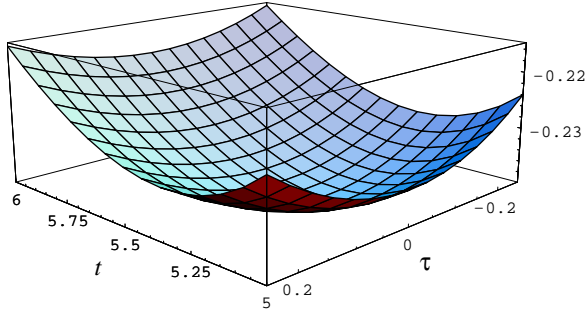}
{\it \caption{The non-supersymmetric saddle point at the field values given
  in (\ref{keepme}), plotted in the $(t,\tau)$ plane.}}
\end{figure}


\subsection{An Example with de-Sitter Turning Points}

Having worked through one example in detail let us briefly give another case.

In the presence of non-perturbative effects it is, of course,
perfectly possible to obtain de Sitter turning points which are
isolated in field space without recourse to any ``raising
mechanism''. To give an example of this let us consider another
example taken from \cite{Micu}.

Retaining the notation in and below \eqref{newvar2}, the table below
summarises the vacua of the heterotic model with superpotential
\begin{equation}\label{sat-W}
  W=8T + 6Z+48 e^{-S/10}.
\end{equation}
As before we only include those ideals which contain
extrema\footnote{The non-supersymmetric vacua with all F-terms
  non-zero is computationally expensive. Restricting to extrema with
  the axions $\tau=\zeta=0$ we find that there are no solutions.}.
\begin{center}
\begin{tabular}{|c|c|c|}\hline
  Ideal & Vacua type & Physical?\\ \hline\hline
  $\langle f_1,f_2,f_3,f_4, f_5,f_6\rangle$ & supersymmetric & 
  Yes, AdS saddles \\ \hline\hline
  $(\gen{\partial V,f_2,f_3,f_4,f_5,f_6}:f_1^\infty)$ & partially F-flat & 
  No, $t=0$ \\\hline
  $(\gen{\partial V,f_1,f_2,f_4,f_5,f_6}:f_3^\infty)$ & partially F-flat & 
  No, $z=0$ \\\hline\hline
  $(((\gen{\partial V,f_2,f_4,f_6}:f^\infty_1):f^\infty_3):f^\infty_5)$ & 
  non SUSY & Yes, dS and AdS saddles \\ \hline\hline
  $(((((\gen{\partial V, f_5}:f^\infty_1):f^\infty_2):f^\infty_3):f^\infty_4):f^\infty_6)$ 
  & non SUSY & No, $z\not\in\mathbb{R}$\\ \hline\hline
\end{tabular}
\end{center} 

As an example let us discuss the ideal, $(((\gen{\partial
  V,f_2,f_4,f_6}:f_1^\infty):f_3^\infty):f_5^\infty)$ in the second to
last row of the above table. We primary decompose this ideal (thus
completing steps 1 and 2) and find \emph{four} primary ideals,
$\mathcal{P}_1\ldots \mathcal{P}_4$. To illustrate, $\mathcal{P}_1$ is
\begin{equation*}\begin{split}
    \langle &y+1,
    \quad 4\tau+3\zeta,\quad 18sx-180xy-20t-15z,\quad 2s^2-45s+50, \\
    &4ts+3zs-117sx+90x,\quad 16tz-12tx-9zx-216x^2, \\
    &64t^2+48tz+36z^2-2412tx-1809zx+18792x^2, \\
    &4z^2s-156zsx+189sx^2+120zx-90x^2, \quad
    16z^3-804z^2x+396tx^2+8649zx^2-20520x^3 \rangle.
\end{split}\end{equation*}
Projecting (step 3) onto the pairs $(\sigma,y)$ and $(s,x)$ we find,
writing rational approximations to the field values valid to the
$10^{-5}$ level,
\begin{equation*}\begin{split}
	0&=y+1\implies \sigma = 10(2n+1)\pi,\quad n\in\mathbb{Z}, \\
	0&=2s^2-45s+50,\implies s_1=\frac{177}{151} \ .
\end{split}\end{equation*}
For the other solution of the quadratic equation in $s$ we find there
are no physical solutions. This eliminates the exponentiated variables
(step 4). We substitute the above values into $V$ and form the new
ideal of derivatives with respect to the remaining variables,
returning us to the perturbative algorithm. The relevant term in the
saturation expansion is a single primary ideal. It is one dimensional
in the variables $\tau,\zeta,t$ and $z$. The physical solutions are
\begin{equation}\begin{split}
	0&=4\tau+3\zeta, \\
	t&= \frac{1089}{151},\qquad z=\frac{1117}{484}, \\
	\text{or }t&=\frac{566}{327},\qquad z=\frac{1452}{151}.
\end{split}\end{equation}
The following table summaries these results and those following from
repeating these steps with the remaining three primary ideals,
$\mathcal{P}_2$, $\mathcal{P}_3$, $\mathcal{P}_4$. None of these
solutions represent isolated vacua - they each have a flat direction
given by $4\tau+3\zeta=0$:
\begin{center}
\begin{tabular}{|c|c|c|c|c|}\hline
  $t$ & $z$ & $s$ & $\sigma$ & Vacuum \\ \hline
  16.590& 22.120& 6.564 &  $10(2n+1)\pi$ & de Sitter saddle,
  $V\simeq 2.10^{-7}$ \\ \hline
  7.212 & 2.308 & 1.172 &  $10(2n+1)\pi$ & AdS saddle, 
  $V\simeq -1.10^{-3}$ \\ \hline
  1.731 & 9.616 & 1.172 &  $10(2n+1)\pi$ & AdS saddle, 
  $V\simeq -1.10^{-3}$ \\ \hline
\end{tabular}
\end{center}

We see then that we do indeed obtain a de Sitter saddle point for this
system. Note that although the value of the classical
cosmological constant for this turning point is of the order of
$10^{-7}$ in the units being used here, and we have approximated the
numerical solutions to the dilaton to order $10^{-5}$, this result is
not a numerical artifact. We can repeat the calculation at greatly
increased precision, say $10^{-12}$ and the values given above
persist.

\begin{center}
\begin{figure}
	\centering\includegraphics[width=0.5\textwidth]{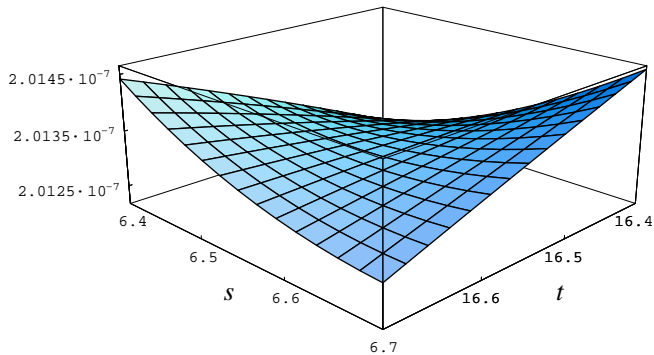}
	{\it \caption{The de Sitter saddle point described in the text,
          plotted in the $s$, $t$ plane. }}
\end{figure}
\end{center}
\subsection{Some IIB Examples}

It is a simple matter, using our methods, to show that the KKLT setup
\cite{Kachru:2003aw},
\begin{equation}\label{kklt}
  W=W_0 + e^{-c T},\qquad K=-3\log(T+\Tb),
\end{equation}
has no vacua other than the well-known supersymmetric AdS minimum. We
define $f_1$ and $f_2$ to be the real and imaginary parts of $F_T$,
and $t$ and $\tau$ to be the real and imaginary parts of $T$. Assuming
$c$ and $W_0$ are non-zero we find that the ideals $(\gen{\partial V,
  f_1}:f_2^\infty)$, $(\gen{\partial V,f_2}:f_1^\infty)$ and $((\gen{\partial V}:f_1^\infty):f_2^\infty)$
each include the generator $ct+2$ which describes an unphysical vacuum at $ct=-2$. This
exhausts the saturation decomposition and thus the supersymmetric vacuum is the only isolated physical extremum.
 
We now consider a slightly more complicated example where the dilaton
is not integrated out.
\begin{equation}
  W=W_0 + a S + e^{-c T},\qquad K=-3\log(T+\Tb)-\log(S+\Sb).
\end{equation}
We shall define, in the rest of this subsection, $f_3$ and $f_4$ to be
the real and imaginary parts of $F_S$ and $s$ and $\sigma$ to be the
real and imaginary parts of $S$.  The following table summarizes the
vacua of this system (as in earlier examples we have assumed some
field values in the computationally expensive final term of the
saturation expansion, as shown in the table).
\begin{center}
\begin{tabular}{|c|c|c|}\hline
  Ideal & Vacua type & Physical Vacua?\\ \hline\hline
  $\langle f_1,f_2,f_3,f_4 \rangle$ & supersymmetric & AdS Saddles \\ 
  \hline\hline
  $((\gen{\partial V,f_2,f_4}:f_1^\infty):f_3^\infty)$ & non SUSY & 
  AdS Saddles\\ \hline
  $((((\gen{\partial V,t-24,\tau}:f_1^\infty)\ldots :f_4^\infty)$ & non SUSY & 
  No\\ \hline
\end{tabular}
\end{center}
Consider primary decomposing the supersymmetric ideal. For simplicity we assume $c$,
$W_0$ and $a$ are real and non-zero and so saturate out the monomial $c W_0 a
t s x y$, which also removes unphysical solutions. We then find two
zero dimensional ideals generated by,
\begin{equation}\label{susyid}
  \langle \sigma,\quad a s - W_0 \mp x,\quad y\mp 1,\quad 
  ctx\pm 3 W_0 + 3x\rangle.
\end{equation}
Here $y=e^{-ic\tau}$ and $x=e^{ -c t}$. Eliminating all of the
variables except $(\tau,y)$ gives the constraint $y\mp 1$ while
eliminating to the pair $(t,x)$ gives the final generator of
(\ref{susyid}).

We must now enumerate the solutions to the transcendental equation
given by the final generator of the ideal \eqref{susyid}.
\bea \label{lambeqn} c t e^{-c t} \pm 3 W_0 + 3 e^{-c t} = 0 \eea This
is simpler than it may at first appear. Equation \eqref{lambeqn} is
solved, in terms of the Lambert W-function $\mathcal{W}_n$, as follows
\cite{Knuth}:
\begin{equation}\label{lam}
	ct=-3-\mathcal{W}_n(\pm 3\,W_0\, e^{-3}) \ .
\end{equation}
We recall the definition of the Lambert W-function as the inverse
function to $f(\mathcal{W}_n) = \mathcal{W}_n \exp( \mathcal{W}_n)$.
Equation \eqref{lam} then represents at most two real solutions for $W_0$
real, given by the branches $n=0$ and $-1$ of the Lambert
W-function. To proceed, let us choose some values for the parameters:
$c=1/10$, and $W_0=-10^{-4}$. The only physical solution in
\eqref{lam} is then obtained when we take the positive sign and when
$n=-1$. We find $t\simeq 107.314$. In addition, from the second
generator in \eqref{susyid}, we have $y=1\implies\tau/10 = 2n\pi$ for
$n\in\mathbb{Z}$.

Having eliminated the variables which appear transcendentally we are
now in a position to return to the perturbative stage of our
method. However, given the simplicity of the ideal \eqref{susyid}, we
may simply read off the remaining field values in this case.
\begin{equation}
	\sigma=0,\quad a s \simeq -7.815\times 10^{-5} \ .
\end{equation}
We see that in order to have a solution at $s>1$ we need a very
small value for $a$. For example, taking $a=-10^{-5}$ we find
$s=7.815$. We then obtain an AdS saddle point, not a minimum,
with cosmological constant $V\simeq -4.5\times 10^{-16}$,
although this turning point is of course Breitenlohner-Freedman
stable.

This system also admits non-supersymmetric vacua. The ideal
$((\gen{\partial V,f_2,f_4}:f_1^\infty):f_3^\infty)$ may be primary
decomposed into four pieces. Two of these pieces require $a=0$ (so
that $S$ contributes only to the K\"ahler potential, not to the
superpotential). The corresponding vacua are not isolated as $\sigma$
is a flat direction. We move on to the case $a\not=0$. There are two
remaining primary ideals, $\mathcal{P}_+$ and $\mathcal{P}_-$, which
we reproduce below:
\begin{equation}\begin{split}
    \mathcal{P}_+ = \langle &y-1,\quad \sigma,\quad 4 c^2 t^2 x+2 a c
    s t+2 c t W_0+8 c t x+7 a s+W_0+x, \\
    &2 a c s t x+3 a^2 s^2-10 c t W_0 x-4 c t x^2+7 a s x-3 W_0^2-5
    W_0 x-2 x^2, \\ &4 a^2 c s^2 t-20 c^2 t^2 W_0 x-8 c^2 t^2 x^2-2 a
    c s t    W_0+6 a c s t x-7 a^2 s^2-6 c t W_0^2 \\
    &\ -10 c t W_0 x-4 c t x^2-a s W_0-a s x, \\
    &6 a^3 s^3-28 a c s t W_0 x-18 a c s t x^2+21 a^2 s^2 x-4 c t
    W_0^2    x-34c t W_0 x^2-12 c t x^3 \\
    &\ -6 a s W_0^2-44 a s W_0 x-17 a s x^2-5 W_0^2 x-7 W_0 x^2-2
    x^3\rangle \ .
\end{split}\end{equation}
\begin{equation}\begin{split}
    \mathcal{P}_- = \langle &y+1,\quad \sigma,\quad 4 c^2 t^2 x-2 a c
    s    t-2 c t W_0+8 c t x-7 a s-W_0+x \\
    &2 a c s t x-3 a^2 s^2-10 c t W_0 x+4 c t x^2+7 a s x+3 W_0^2-5
    W_0 x+2 x^2, \\ &4 a^2 c s^2 t+20 c^2 t^2 W_0 x-8 c^2 t^2 x^2-2
    a c s t W_0-6 a c s t x-7 a^2 s^2-6 c t W_0^2 \\
    &\ +10 c t W_0 x-4 c t x^2-a s W_0+a s x, \\
    &6 a^3 s^3+28 a c s t W_0 x-18 a c s t x^2-21 a^2 s^2 x+4 c t
    W_0^2    x-34 c t W_0 x^2+12 c t x^3 \\
    &\ -6 a s W_0^2+44 a s W_0 x-17 a s x^2+5 W_0^2 x-7 W_0 x^2+2
    x^3\rangle \ .
\end{split}\end{equation}

Consider first the ideal $\mathcal{P}_+$. Projecting the corresponding
variety onto the $(\tau,y)$ plane we find $y=1$. Projecting onto the
$(t,x)$ plane and using the definition of $x$ we find the following
equation for $t$ \bea f_+(t) &\equiv& 4 c^4 t^4 e^{-2 c t}+4 c^3 t^3
e^{-2 c t}-30 c^2 t^2
W_0 e^{-c t}-41 c^2 t^2 e^{-2 c t}-9 c t W_0^2-87 c t W_0 e^{-c t} \\
\nonumber &&-78 c t e^{-2 c t}-18 W_0^2-36 W_0 e^{- c t}-18 e^{-2 c
  t}=0.  \eea

Performing a similar analysis for $\mathcal{P}_-$, we find $y=-1$ and
\bea f_-(t) &\equiv& 4 c^4 t^4 e^{-2 c t}+4 c^3 t^3 e^{-2 c t}+30 c^2
t^2 W_0 e^{-c t}-41 c^2 t^2 e^{-2 c t}-9 c t W_0^2+87 c t W_0 e^{- c
  t} \\ \nonumber && -78 c t e^{-2 c t}-18 W_0^2+36 W_0 e^{-c t}-18
e^{-2 c t}=0.  \eea

Let us take the same values of $c$, $W_0$ and $a$ ($1/10$, $- 10^{-4}$
and $- 10^{-5}$ respectively) as for the above analysis of the ideal
of supersymmetric vacua. Solving these equations requires use of the
numerical step of our algorithm, for which we employ Mathematica. We
will look for vacua with $1\leq t \leq 250$ and we start by
enumerating the number of vacua in this range. Beginning with
$\mathcal{P}_+$, we perform the contour integration described in
Section \ref{generalmethod},
\begin{equation}
	\oint \!\ud z\,\, \frac{\partial_z f_+(z)}{f_+(z)},
\end{equation}
with the `radius' of the contour initially set at $0.1$, so $z=t-0.1i$
on the straight line segment lying below the real axis, for
example. This calculation tells us that there are two roots. Having
found the number of roots we now use a numerical root finding routine
for uni-variate systems to locate them. One root is located at
$t_1\simeq 35.396\ldots$. We now perform an integral over the circular
contour $z=35.396+0.1\exp i\theta$ and find the multiplicity of this
root to be one. The second and final root is found at $t_2\simeq
133.411$, and is also of multiplicity one. The ideal $\mathcal{P}_-$
yields two further roots on following the same procedure. We are now
certain to have found all of the solutions for $t$ in the specified
range.

Once we have eliminated the transcendental variables in this manner we
return to the perturbative algorithm. We find that only the values
$y=1$ and $t=t_1$ (i.e. one of the roots of the ideal $\mathcal{P}_+$)
lead to a genuine physical vacuum, which is an AdS saddle point at
$\sigma=0$, $s\simeq 16.369$ with $V\simeq -3.6\times 10^{-13}$. This
vacuum is not Breitenlohner-Freedman stable.

\begin{figure}	
\centering\includegraphics[width=0.5\textwidth]{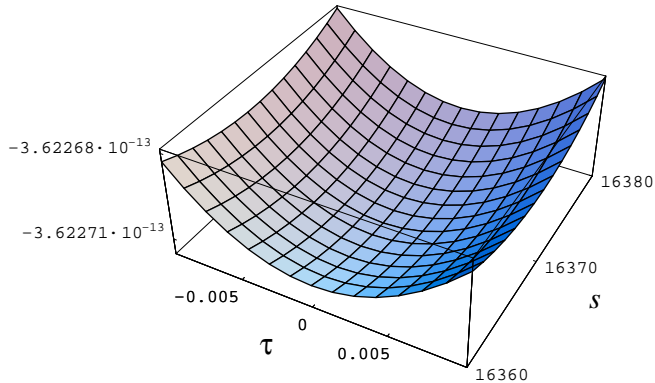}
{\it \caption{The IIB AdS saddle point at $\tau=0$, $t\simeq35.396$,
  $s\simeq 16,369$, $\sigma=0$ plotted in the $\tau,s$ plane}}
\end{figure}


\section{Constraints on Flux Parameters}
In addition to finding vacua, we can also place powerful constraints
on the parameters that arise in flux compactifications. This was done
in \cite{Gray:2006gn} for the case of perturbative
superpotentials\footnote{Alternative methods for finding constraints in the
case of non-supersymmetric Minkowski vacua can be found, for example,
in \cite{Gomez-Reino:2006dk,Gomez-Reino:2006wv,Soroush:2007ed}.}. The essential tool
here is the elimination of variables in an ideal, which we have
already discussed in the context of projection onto the space spanned
by the dummy and exponentiated variables in the previous section.

Given some K\"ahler and superpotential we take the set of parameters
which are present, call them $a_{\alpha}$, as variables and consider
our initial ideal, $\gen{ \partial V }$ or part of its saturation
decomposition, as living in the ring
$\mathbb{C}[m_i,\mu_i,n_a,\nu_a,x_a,y_a,a_{\alpha}]$. We then
eliminate the field variables
$m_i,\mu_i,n_a,\nu_a,x_a,y_a$. Algebraically this is the intersection
of our ideal with the ring $\mathbb{C}[m_i,\mu_i,n_a,\nu_a,x_a,y_a]$.
Geometrically, this is the projection of our ideal onto the surface
where the field variables and their exponentials vanish. The elements
of the resulting ideal then represent necessary constraints on the
parameters for extrema to exist. These, of course, apply in addition
to any constraint already imposed on the system by other
considerations. 

As a simple example, consider the following K\"ahler and
superpotentials obtained from the heterotic string compactified on
generalised half-flat manifolds in the presence of flux and gaugino
condensation \cite{Micu}:
\begin{equation}\begin{split}\label{cons-eg}
    W&=i(\xi+ieT)+(\epsilon +ipT)Z+\frac{i}{2}(\mu+iqT)Z^2 +\lambda
    e^{-cS}, \\
    K&=-3\log(T+\Tb) -3\log(Z+\bar{Z}) -\log(S+\Sb) \ .
\end{split}\end{equation} 
In the notation of our general discussion we would write $ \{ M_i= m_i
+ i \mu_i \} = \{ Z= z + i \zeta,T= t + i \tau \} $ and $ \{ N = n + i \nu
\} = \{ S= s + i \sigma \} $.

In this system, the flux parameters obey the constraint,
\begin{equation}
\label{knownconstr}
	\epsilon q-\mu p =0,
\end{equation}
which follows from the Bianchi identity in the case of a `standard
embedding' \cite{Micu}.

To demonstrate, let us consider the supersymmetric vacua of this
system. The properties of the supersymmetric vacua are determined by
the ideal generated by the real and imaginary parts of the F-terms and
the constraint \eqref{knownconstr}, $\gen{
  f_1,f_2,f_3,f_4,f_5,f_6,-\epsilon q+\mu p}$, in the ring
$\mathbb{C}[\xi,e,\epsilon,p,\mu,q,\lambda,c,s,t,z,\sigma,\tau,\zeta]$. Here
$f_1$ and $f_2$ are the real and imaginary parts of $F_T$, $f_3$ and
$f_4$ the real and imaginary parts of $F_Z$ and $f_5$ and $f_6$ the
real and imaginary parts of $F_S$. In addition we are making our
habitual complexification of the field, and now parameter, space.

One question we could ask is, given some physical constraints that we
wish the field values to obey in vacuum, what constraints are
necessary amongst the parameters for the existence of a supersymmetric
vacuum of that form? Looking at the strengths of the gravitational and
gauge interactions, and insisting on a vanishing theta term for the
gauge fields, one might choose a set of field values of the form
$s=24,t=1,\sigma=0$.  These physical requirements can be enforced by
including appropriate extra generators in our initial ideal. In this
case, this would correspond to taking the ideal $\gen{
  f_1,f_2,f_3,f_4,f_5,f_6,-\epsilon q + \mu
  p,s-24,t-1,\sigma }$. We then eliminate all of the field variables
using a Gr\"obner basis calculation as described above. The result is
the following set of constraints amongst the parameters.

\beq
\ba{rcl}
0 &=& \mu p - q \epsilon \\ 
0 &=& p q \xi - e q \epsilon \\
0 &=& p^2 \xi - e p \epsilon \\
0 &=& 96 c p \xi - 96 c e \epsilon - p \xi + e \epsilon \\
0 &=& e \mu q - q^2 \xi \\
0 &=& e^2 \mu - e q \xi \\
0 &=& 96 c e \mu - 96 c q \xi - e \mu + q \xi \ .
\ea
\label{constrideal}\eeq

Thus, if we are looking for supersymmetric vacua in compactifications
of Heterotic on generalised half-flat manifolds with gaugino condensation
in which the interaction strengths are physical and the theta angle is
zero, then the above constraints among the parameters in $W$ must hold.

We can in fact do even better. By using various pieces of the
methodology espoused in this paper, and in \cite{Gray:2006gn}, we can
make other demands of the system. For example, we may require that
none of the parameters are zero (so that all terms in (\ref{cons-eg})
contribute). To achieve this we simply take the ideal generated by the
right hand sides in \eqref{constrideal} and saturate out the
polynomial which is the product of the parameters, $e c \mu \xi p q
\epsilon$. Given the geometrical interpretation of saturation, as
discussed in Section \ref{generalmethod}, this then gives us equations
for the parts of the variety of constraints in parameter space for
which none of the parameters vanish.

\beq\ba{rcl}
0 &=& p \xi - e \epsilon \\
0 &=& \mu p - q \epsilon \\
0 &=& e \mu - q \xi \ .
\ea\eeq

One of these three constraints is redundant and another is simply that
which follows from the Bianchi identity
\eqref{knownconstr}. This thus leaves us with a single additional
simple constraint on the parameters, $e \mu - q \xi = 0$.

Such constraints are clearly of great utility in searching for vacua
of these systems - whether using the methods presented in this paper
or more conventional techniques. A quick Gr\"obner basis calculation
can cut out huge swathes from the parameter space which needs to be
searched.

\section{Conclusions}

We have shown that the algebro-geometric techniques, advocated in
\cite{Gray:2006gn}, for finding vacua of supergravity systems can be
modified to incorporate the effects of non-perturbative contributions
to the superpotential. While the modification required does introduce
a numerical element to our method, much of the information that can be
obtained remains analytic. In particular, we can analytically prove
that we do not miss any vacua due to this numerical step.

While not a universal cure to the problems of finding vacua in string
phenomenology systems, these methods do render the process
significantly less time consuming and painful. Many of the vacua which
spontaneously break supersymmetry and may be found with our methods
are very unlikely to be found with more conventional techniques. In
many cases the techniques espoused here represent the most powerful
tool currently available in searching for vacua in a given theory.

The algorithmic nature of the methods presented in this paper and in
\cite{Gray:2006gn} is important. It means that it is conceivable to
create a `black box' package to which one feeds information such as a
K\"ahler and Superpotential, ranges for the parameters of interest and
physical requirements on the vacua to be found. The program would then
simply cycle through the specified parameter range, outputting any
vacua present of the type desired. A user specified time constraint
would impose an upper limit on how far along the saturation expansion
one could reach in each case. It is the intention of the authors to
make such a black box available as the next step in this program of
research.

\section*{Acknowledgements}
J.~G.~is supported by CNRS. Y.-H.~H.~is supported by the FitzJames
Fellowship of Merton College, Oxford.
A.~I.~is grateful for support from the 
European network EUCLID (HPRN-CT-2002-00325) during this project. A.~L.~is supported by
the EC 6th Framework Programme MRTN-CT-2004-503369.


\begin{thebibliography}{99}

\bibitem{Giddings:2001yu}
  S.~B.~Giddings, S.~Kachru and J.~Polchinski,
  ``Hierarchies from fluxes in string compactifications,''
  Phys.\ Rev.\ D {\bf 66} (2002) 106006
  [arXiv:hep-th/0105097].

\bibitem{Kachru:2003aw}
  S.~Kachru, R.~Kallosh, A.~Linde and S.~P.~Trivedi,
  ``De Sitter vacua in string theory,''
  Phys.\ Rev.\ D {\bf 68} (2003) 046005
  [arXiv:hep-th/0301240].

\bibitem{Balasubramanian:2005zx}
  V.~Balasubramanian, P.~Berglund, J.~P.~Conlon and F.~Quevedo,
  ``Systematics of moduli stabilisation in Calabi-Yau flux
  compactifications,''
  JHEP {\bf 0503}, 007 (2005)
  [arXiv:hep-th/0502058].

\bibitem{Grana:2005jc}
  M.~Grana,
  ``Flux compactifications in string theory: A comprehensive review,''
  Phys.\ Rept.\  {\bf 423}, 91 (2006)
  [arXiv:hep-th/0509003].

\bibitem{Buchbinder:2003pi}
  E.~I.~Buchbinder and B.~A.~Ovrut,
  ``Vacuum stability in heterotic M-theory,''
  Phys.\ Rev.\  D {\bf 69}, 086010 (2004)
  [arXiv:hep-th/0310112].

\bibitem{Gray:2007mg}
  J.~Gray, A.~Lukas and B.~Ovrut,
  ``Perturbative anti-brane potentials in heterotic M-theory,''
  arXiv:hep-th/0701025.

\bibitem{deCarlos:2004ci}
  B.~de Carlos, A.~Lukas and S.~Morris,
  ``Non-perturbative vacua for M-theory on G(2) manifolds,''
  JHEP {\bf 0412}, 018 (2004)
  [arXiv:hep-th/0409255].

\bibitem{Acharya:2006ia}
  B.~Acharya, K.~Bobkov, G.~Kane, P.~Kumar and D.~Vaman,
  ``An M theory solution to the hierarchy problem,''
  Phys.\ Rev.\ Lett.\  {\bf 97}, 191601 (2006)
  [arXiv:hep-th/0606262].

\bibitem{Acharya:2007rc}
  B.~S.~Acharya, K.~Bobkov, G.~L.~Kane, P.~Kumar and J.~Shao,
  ``Explaining the electroweak scale and stabilizing moduli in M theory,''
  arXiv:hep-th/0701034.

\bibitem{Micu:2007rd}
  A.~Micu, E.~Palti and G.~Tasinato,
  ``Towards Minkowski vacua in type II string compactifications,''
  arXiv:hep-th/0701173.

\bibitem{Gray:2006gn} 
  J.~Gray, Y.~H.~He and A.~Lukas, 
  ``Algorithmic algebraic geometry and flux vacua,'' 
  JHEP {\bf 0609}, 031 (2006) 
  [arXiv:hep-th/0606122]. 

\bibitem{probe} 
   J.~Gray, Y.~H.~He, V.~Jejjala and B.~D.~Nelson,
  ``Exploring the vacuum geometry of N = 1 gauge theories,''
  Nucl.\ Phys.\  B {\bf 750}, 1 (2006)
  [arXiv:hep-th/0604208].\\ 
   J.~Gray, Y.~H.~He, V.~Jejjala and B.~D.~Nelson,
  ``Vacuum geometry and the search for new physics,''
  Phys.\ Lett.\  B {\bf 638}, 253 (2006)
  [arXiv:hep-th/0511062].

\bibitem{Distler:2005hi}
  J.~Distler and U.~Varadarajan,
  ``Random polynomials and the friendly landscape,''
  arXiv:hep-th/0507090.

\bibitem{Strominger:1995cz}
  A.~Strominger,
  ``Massless black holes and conifolds in string theory,''
  Nucl.\ Phys.\  B {\bf 451}, 96 (1995)
  [arXiv:hep-th/9504090].

\bibitem{mac} 
D.~Grayson and M.~Stillman, 
``Macaulay 2, a software system for research in algebraic geometry.'' 
Available at {\tt http://www.math.uiuc.edu/Macaulay2/} 
 
\bibitem{sing} 
G.-M.~Greuel, G.~Pfister, and H.~Sch\"onemann, 
''Singular: A Computer Algebra System for Polynomial Computations,'' 
Centre for Computer Algebra, University of Kaiserslautern (2001). 
Available at {\tt http://www.singular.uni-kl.de/} 
 
\bibitem{Breitenlohner:1982bm} 
  P.~Breitenlohner and D.~Z.~Freedman, 
  ``Positive Energy In Anti-De Sitter Backgrounds And Gauged Extended 
  Supergravity,'' 
  Phys.\ Lett.\ B {\bf 115}, 197 (1982).

\bibitem{mathematica}
Mathematica Version 5.2, Wolfram Research, Inc. 2005.

\bibitem{maple}
Maple 10.06, Waterloo Maple Inc. (Maplesoft). 2006.

\bibitem{Micu}
  S.~Gurrieri, A.~Lukas and A.~Micu,
  ``Heterotic on half-flat,''
  Phys.\ Rev.\ D {\bf 70}, 126009 (2004)
  [arXiv:hep-th/0408121].\\
 B.~de Carlos, S.~Gurrieri, A.~Lukas and A.~Micu,
  ``Moduli stabilisation in heterotic string compactifications,''
  JHEP {\bf 0603}, 005 (2006)
  [arXiv:hep-th/0507173].

\bibitem{Knuth} R.~M.~Corless et al, ``On the Lambert W function''
  Adv.\ Comp.\ Maths {\bf 5} (1996) 329

\bibitem{Gomez-Reino:2006dk}
  M.~Gomez-Reino and C.~A.~Scrucca,
  ``Locally stable non-supersymmetric Minkowski vacua in supergravity,''
  JHEP {\bf 0605} (2006) 015
  [arXiv:hep-th/0602246].  

\bibitem{Gomez-Reino:2006wv}
  M.~Gomez-Reino and C.~A.~Scrucca,
  ``Constraints for the existence of flat and stable non-supersymmetric vacua
  in supergravity,''
  JHEP {\bf 0609} (2006) 008
  [arXiv:hep-th/0606273].

\bibitem{Soroush:2007ed}
  M.~Soroush,
  ``Constraints on meta-stable de Sitter flux vacua,''
  [arXiv:hep-th/0702204].

\end{thebibliography}
\end{document}